\documentclass[10pt, twocolumn, a4paper, superscriptaddress]{revtex4-2}

\usepackage{amsmath}
\usepackage{graphicx}
\usepackage{color}
\usepackage{hyperref}

\begin{document}

\title{Higher-order topological phase induced by hybrid magneto-electric resonances}

\author{Daniel A. Bobylev}
\thanks{These two authors contributed equally to this work}
\affiliation{School of Physics and Engineering, ITMO University, Saint Petersburg, Russia}
\author{Dmitry V. Zhirihin}
\thanks{These two authors contributed equally to this work}
\affiliation{School of Physics and Engineering, ITMO University, Saint Petersburg, Russia}
\author{Dmitry I. Tihonenko}
\affiliation{School of Physics and Engineering, ITMO University, Saint Petersburg, Russia}
\author{Anton Vakulenko}
\affiliation{Department of Electrical Engineering, Grove School of Engineering, City College of the City University of New York, New York, NY, USA}
\affiliation{Physics Program, Graduate Center of the City University of New York, New York, NY, USA}
\author{Daria A. Smirnova}
\affiliation{Research School of Physics, Australian National University, Canberra, ACT, Australia}
\affiliation{Institute of Applied Physics, Russian Academy of Science, Nizhny Novgorod, Russia}
\author{Alexander B. Khanikaev}
\affiliation{Department of Electrical Engineering, Grove School of Engineering, City College of the City University of New York, New York, NY, USA}
\affiliation{Physics Program, Graduate Center of the City University of New York, New York, NY, USA}
\author{Maxim A. Gorlach}
\affiliation{School of Physics and Engineering, ITMO University, Saint Petersburg, Russia}


\maketitle

{\bf Rapid development of topological concepts in photonics unveils exotic phenomena such as unidirectional propagation of electromagnetic waves resilient to backscattering at sharp bends and disorder-immune localization of light at stable frequencies~\cite{Lu2014,Ozawa}. Recently introduced higher-order topological insulators (HOTIs)~\cite{BenalcazarScience,Schindler2018} bring in additional degrees of control over light confinement and steering. However, designs of photonic HOTIs reported so far are solely exploiting lattice geometries which are hard to reconfigure thus limiting tunability. Here, we elaborate a conceptually new mechanism to engineer higher-order topological phases which relies on the dual nature of  electromagnetic field and exploits both electric and magnetic responses of the meta-atoms. Hybridization between these responses gives rise to the difference in the effective coupling which is controlled by the meta-atoms mutual orientations. This feature facilitates us to tailor photonic band topology exclusively via particle alignment and to flexibly reconfigure the topological phase. Focusing on the kagome array of split-ring resonators, we experimentally demonstrate topological edge and corner states in the microwave domain. Our findings provide a new promising route to induce and control higher-order topological phases and states.}


The first approaches to orchestrate photonic topological states were inspired by condensed matter physics and largely provided photonic analogues of quantum Hall~\cite{RaghuHaldane2008,Wang2009,Skirlo2015} and quantum spin Hall~\cite{Hafezi2013,Mittal2014,KhanikaevNatMat} effects. Further investigations have revealed such unique advantages of photonic systems as possibility to arbitrarily tailor lattice geometry~\cite{WuHuPRL}, introduce non-Hermiticity~\cite{Weimann2016,Szameit2021}, combine electric and magnetic responses of the constituents~\cite{KhanikaevNatMat,Slob2016,SlobNPhot,Slob2019,Bobylev2020},  and leverage nonlinear optical phenomena~\cite{Lumer2013,Leykam2016,NonlTopPhot,NonlinearRechtsman,Kirsch2021}.

Higher-order topological states (HOTS) signaled a new twist in topological physics and enabled resilient light localization in photonic structures of different dimensionality~\cite{BenalcazarScience,Schindler2018}. Shortly after their prediction~\cite{BenalcazarScience}, higher-order topological states were realized in a variety of systems across the entire electromagnetic spectrum from radiofrequencies~\cite{Imhof2018} and microwaves~\cite{Peterson2018,PhotonicKagome} to optical wavelengths~\cite{Noh2018,Mittal2019,Hassan2019,Vakulenko2021}.

The designs of photonic HOTIs reported so far largely rely on the geometric lattice symmetries which provide a crucial ingredient in the formation of topological phases. However, because lattice geometry is difficult to reconfigure, these realizations cannot offer desirable tunability. 

In this Letter, we put forward an alternative mechanism to engineer higher-order topological phases and states. Our approach relies on the possibility of a versatile control of interactions unique to electromagnetic systems endowed with both electric and magnetic degrees of freedom which can be controllably mixed via magneto-electric (bianisotropic) response. This enables a novel platform for higher-order topological photonics where topological properties are induced not through the control of the lattice symmetries and inter-particle distances, but rather through the mutual orientation of the meta-atoms. To confirm the proposed concept, we investigate two structures based on the same lattice but formed by the differently orientated bianisotropic split-ring resonators~\cite{Pendry1999}. We show that the two systems are characterized by the distinct topological invariants. As a result, the boundaries between them host topologically protected states localized at the edges and corners. We also confirm our conclusions experimentally in the microwave domain.


\begin{figure*}[t]
    \centering
    \includegraphics{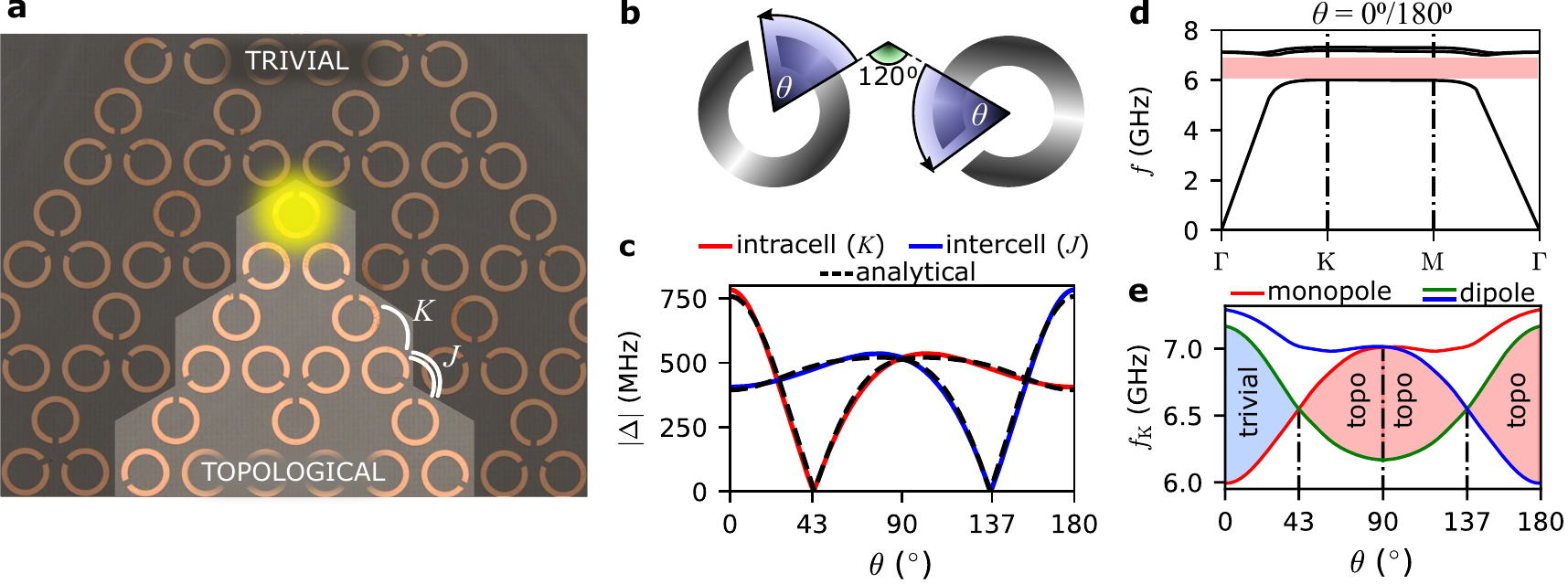}
    \caption{{\bf Designing band topology via mutual orientation of bianistropic split-ring resonators (SRRs) arranged in kagome lattice.} (a) The sketch of the designed structure based on kagome lattice of bianisotropic SRRs. The trivial domain with inward gap orientation is shaded. The position of the topological corner state is highlighted in yellow. SRR rotation angle $\theta$ is specified in panel (b).  (b) Relative orientation of two SRRs comprising the designed structure. Each of the meta-atoms is rotated by the angle $\theta$ counter-clockwise. (c) Absolute values of effective intracell and intercell couplings of the meta-atoms versus rotation angle determined from full-wave numerical simulations of eigenfrequency splitting in Comsol Multiphysics. Dashed line shows the fitting of numerical results by Eq.~\eqref{eq:splitting}. (d) Band diagram calculated for the lowest-frequency hybrid dipole modes supported by the designed structure. Complete photonic bandgap is shown by light red. (e) The frequencies of the lowest-frequency dipole modes in K point of the Brillouin zone versus rotation angle $\theta$. Band inversions indicate the topological transitions happening in the system.}
     \label{fig:KagomeDesign}
\end{figure*}


{\it Proposed design and theoretical model~--~} To illustrate our approach to higher-order topological phase engineering, we design the structure based on the kagome array of split-ring resonators (SRRs) depicted in Fig.~\ref{fig:KagomeDesign}{\bf a}. The individual SRR [Supplementary Materials, Fig.~S1{\bf a}] supports three low-frequency modes one of which is pure electric dipole, whereas the remaining two exhibit dual nature possessing both in-plane electric and out-of-plane magnetic dipole moments. Such hybrid modes expand the landscape of possible interactions within the lattice and enable additional control over coupling between the meta-atoms through their mutual orientation.

To understand the nature of interactions between SRRs in the lattice, we consider a pair of meta-atoms shown in Fig.~\ref{fig:KagomeDesign}{\bf b} tracing the splitting of the lowest-frequency hybrid magneto-electric mode. Due to interaction between SRRs, this mode splits into two states with the frequencies $f_s$ and $f_a$ having parallel and anti-parallel magnetic moments, respectively. Since the eigenfrequency splitting $\Delta(\theta) = f_{a}-f_{s}$ is an even periodic function of SRR's rotation angle $\theta$ [Fig.~\ref{fig:KagomeDesign}{\bf b}], it can be presented in the form of the Fourier series
\begin{equation}\label{eq:splitting}
\Delta(\theta) = \sum_{n} A_n \cos n\theta,
\end{equation}
where $n=0, 1, 2, \dots$ and the coefficients $A_n$ can be calculated theoretically or extracted from the full-wave numerical simulations (Supplementary Note~1). The number of non-negligible terms necessary to capture the interaction between the meta-atoms depends on the distance between them being equal to 5 for our particular design [see parameters in the Methods section]. The fit of numerical results by Eq.~\eqref{eq:splitting} is provided in Fig.~\ref{fig:KagomeDesign}\textbf{c}.

Harmonics $\cos n\theta$ with odd $n$ describe the difference between the splittings $\Delta(\theta)$ and $\Delta(\pi-\theta)$, which opens a way to engineer the arrays with alternating coupling patterns. Remarkably, the splitting of the hybrid mode exhibits strong dependence on the angle $\theta$. For instance, $|\Delta(0^{\circ})| \approx 780$~MHz, whereas $|\Delta(180^{\circ})| \approx 400$~MHz, yielding the ratio of the two coupling constants $\xi \approx |\Delta(0^{\circ})/\Delta(180^{\circ})| \approx 2$. Furthermore, this geometry allows sign reversal of the frequency splitting $\Delta$ which in our case happens for the angles $\theta$ equal to $43^{\circ}$ and $137^\circ$.

{\it Mode dispersion and topological invariant~--~} The discussed orientation-dependent coupling of bianistropic meta-atoms has a profound effect on band topology. If the gaps of split-ring resonators are pointing towards the center of the unit cell, as in the outer domain in Fig.~\ref{fig:KagomeDesign}{\bf a}, the rotation angle $\theta=0^\circ$ and hence the coupling between the meta-atoms within the unit cell (intracell) is strong compared to that between the neighboring unit cells (intercell coupling), see Fig.~\ref{fig:KagomeDesign}{\bf c}. This scenario mimics topologically trivial shrunken kagome lattice~\cite{Ezawa2018,Xue2018,Ni_2018}. On the contrary, the outward orientation of the SRRs gaps, as in the inner domain in  Fig.~\ref{fig:KagomeDesign}{\bf a} ($\theta=180^\circ$) renders weaker intracell coupling compared to the intercell one. This situation is equivalent to the case of topologically nontrivial expanded kagome lattice. 

At the same time, both of the described domains share an identical bulk band structure, since one differs from another solely by the unit cell choice. The calculated dispersion of the three lowest-frequency dipole bands is presented in Fig.~\ref{fig:KagomeDesign}{\bf d} featuring a complete photonic bandgap. The spectral width of this bandgap is strongly sensitive to the SRR rotation angle $\theta$ vanishing for some angles and exhibiting a band inversion clearly seen in Fig.~\ref{fig:KagomeDesign}{\bf e}.

\begin{figure*}[t]
    \centering
    \includegraphics[width=\linewidth]{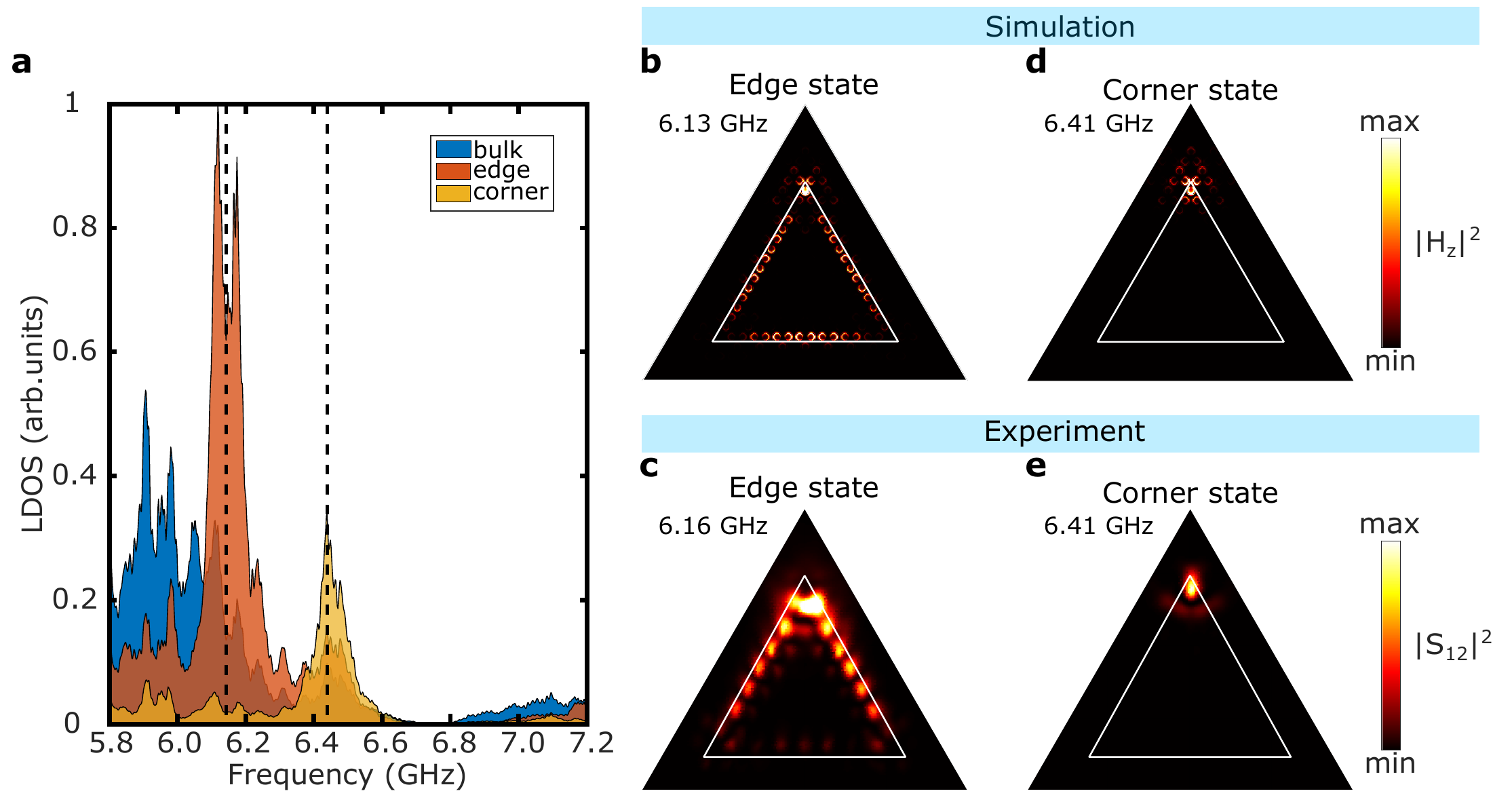}
    \caption{{\bf Experimental investigation of topological edge and corner states.} (a) The retrieved local density of states (LDOS) at the corners, edge and in the bulk of the structure. The structure is excited by electric dipole antenna placed at the corner. Characteristic peaks occur at the frequencies of the respective modes. Black dashed lines mark the spectral positions of the peaks obtained from numerical simulations. (b-e) Field distributions corresponding to the excitation of the edge (b,c) and corner (d,e) mode obtained from full-wave numerical simulations in CST Microwave Studio (b,d) and experiment (c,e). Color encodes  $|H_z|^2$ and $|S_{12}|^2$ for simulations and experiment, respectively.}
    \label{fig:Experiment}
\end{figure*}

Thus, the designed array of rotated SRRs behaves similarly to the breathing kagome lattice known to host topological corner states~\cite{Ezawa2018,Xue2018,Ni_2018}. However, the topology of the bands is governed here not by the interparticle distance as in the previous realizations but through the spatially varying anisotropic  magneto-electric coupling controlled via meta-atoms orientation. The band inversion, which occurs for the certain rotation angles, indicates a topological phase transition in the system.



To confirm our predictions, we simulate the periodic array of split-ring resonators for the two possible unit cell choices and retrieve the Bloch modes formed by the dipole moments of the meta-atoms at the $\Gamma$ and {K} points of the first Brillouin zone. By inspecting their symmetry with respect to the $C_3$ rotation we determine the topological invariant~\cite{SymmetryInvariants} (Supplementary Note~2).
Our results yield nonzero invariant $\chi^{(3)} = (-1, 0)$ for the unit cell where the gaps of the SRRs are pointing outwards the center, while SRR  structure with inward gap orientation appears to be topologically trivial with $\chi^{(3)} = (0, 0)$, which fully agrees with the picture of effective coupling links in the breathing kagome lattice. As a consequence, the boundaries of the two domains [Fig.~\ref{fig:KagomeDesign}\textbf{a}] are expected to host topological edge and corner states.

{\it Experimental observation of topological edge and corner states~--~} To test our predictions, we have fabricated a structure composed of copper split-ring resonators printed on a dielectric substrate as further described in Methods. Leveraging the dual nature of SRRs, we have excited the structure by electric dipole antenna placed on the upper side of the printed circuit board (PCB) and measured $S_{12}$ transmission coefficient through the sample by the loop antenna located on the opposite side of PCB. The measurements performed in the range from 5.8 to 7.2~GHz allowed us to map the distribution of $H_z$ component of the near field at various frequencies.

When the frequency of the source matches the frequency of the mode, field intensity increases exhibiting a series of characteristic peaks depicted in Fig.~\ref{fig:Experiment}{\bf a}. To distinguish the  different types of modes, we employ spatial filtering technique showing the overall intensity of the near field (Fig.~\ref{fig:Experiment}{\bf a}, blue color), field intensity from the edge SRRs (Fig.~\ref{fig:Experiment}{\bf a}, red color) and from the corner SRRs (Fig.~\ref{fig:Experiment}{\bf a}, orange color) as further detailed in Supplementary Note~3. Hence, characteristic maxima in the frequency dependence of the retrieved local densities of states correspond to the bulk, edge and corner states depending on the chosen spatial filter. By examining the near-field profiles at the respective frequencies, we observe edge- and corner-localized field distributions (Fig.~\ref{fig:Experiment}{\bf c,e})  evidencing the emergence of topological edge and corner states, which is in agreement with the results of the full-wave numerical simulations (Fig.~\ref{fig:Experiment}{\bf b,d}).

Interestingly, due to the presence of long-range interactions in our system,  we have  also revealed more subtle type-II corner states~\cite{PhotonicKagome} absent in the conventional nearest-neighbor coupled  breathing kagome lattices and splitting from the band of edge states as further discussed in Supplementary Note 3.


{\it Discussion~--~} To summarize, we have put forward a novel mechanism to design and reconfigure photonic topological phases by exploiting hybrid magneto-electric resonances of bianistropic meta-atoms. While the developed strategy was demonstrated on the example of kagome lattice of split-ring resonators, it can be readily generalized to create quadrupole topological insulators as well as three-dimensional topological structures thus opening a promising direction towards dynamically reconfigurable domain walls and topological cavities. In a broader perspective, our approach enables resilient higher-order topological structures with subwavelength localization which can serve as a compact and versatile platform for future topological meta-devices.

\small{
\section*{Methods}

\textbf{Sample fabrication.} The experimental sample was fabricated using Printed Circuit Board (PCB) technology on Arlon 255C substrate with the relative permittivity 2.55. Geometric sizes of the substrate are $220\times 190\times 1$~mm. Printed structure consists of the two domains of metallic split-ring resonators (SRRs) with the different gap orientations which defined the topological properties. An inner topologically nontrivial domain had a triangular shape with 8 unit cells per triangle side. An outer domain had the thickness of 2 unit cells which according to our simulations is sufficient to ensure the decay of topological edge and corner states.

Parameters of the unit cell of the fabricated structure are as follows: inner radius of the ring $r= 2.25$~mm; ring width $t$ and ring gap $b$ are both equal to 0.75~mm, lattice constant $a=15$~mm. To manipulate the magnitude of the effective couplings, the gaps of the rings were turned towards the unit cell center in the outer trivial domain and outside the center in the inner topological domain.

\textbf{Dependence of the eigenfrequency splitting on SRR rotation angle.} The splitting of the lower-frequency hybrid mode for a pair of interacting SRRs is fitted by the equation $\Delta(\theta) = \sum_{n} A_n \cos n\theta$. For the designed structure, the values of the coefficients read: $A_0 = -210.9$ MHz, $A_1 = 453.6$ MHz, $A_2 = 347.8$ MHz, $A_3 = 122.4$ MHz and $A_4 = 44.7$ MHz, while the contribution of higher harmonics can be neglected.

\textbf{Measurement of the near fields.} To reveal topological edge and corner states supported by the structure, we have measured the map of the transmission coefficient $S_{12}$ over the entire sample in the frequency range $5.8-7.2$~GHz. The transmission was measured between the two antennas placed from the opposite sides of the PCB. To excite the structure, we have used an in-plane electric dipole antenna placed close to the corner SRR of the inner domain. As a receiving antenna, we have used magnetic probe Langer SX-R 3-1 with a loop parallel to the PCB attached to the three-axis precision scanner. Hence, the obtained result quantifies the magnitude of $H_z$ component of magnetic field in the vicinity of the sample. The transmission coefficient has been measured with the help of vector network analyzer Rohde \& Schwarz ZVB20.

Post-processing the obtained maps of $S_{12}$ coefficient, we have also extracted local density of states (LDOS) which is calculated as a sum of squared amplitudes of $S_{12}$ parameter multiplied by the Heaviside step function taking nonzero values only in the specific parts of the sample which include either bulk or edge or corner sites as defined by the spatial filter function (see Supplementary Note~3). Clear maxima in the obtained frequency dependence of LDOS indicate the characteristic frequencies of bulk, edge and corner states. The field maps presented in Fig.~\ref{fig:Experiment}{\bf b-e} correspond to the frequencies of those peaks.
}

\section*{Acknowledgments}
This work was supported by the Russian Science Foundation (grant No.~21-79-10209). D.A.S. acknowledges support from the Australian Research Council (DE190100430).

\section*{Author contributions}
A.B.K. and M.A.G. conceived the idea. D.A.B. and M.A.G. elaborated the theoretical model. D.A.B., D.I.T. and D.A.S. performed numerical simulations. D.V.Z., D.I.T. and A.V. designed the experimental sample. D.V.Z. and D.I.T. performed the experimental measurements. M.A.G. supervised the project. D.A.B. and M.A.G. prepared the manuscript with the input from all the authors.

\section*{Data availability}
The data that support the findings of this study are available from the corresponding author upon reasonable request.

\section*{Competing interests}
The authors declare that they have no competing interests.

\section*{Additional information}
Correspondence and requests for materials should be addressed to M.A.G. (email: m.gorlach@metalab.ifmo.ru).

\bibliography{references}

\end{document}


\maketitle

\newpage

\tableofcontents

\newpage

\section{Supplementary Note 1 -- Derivation of orientation-dependent coupling for bianisotropic meta-atoms}
\label{sec:Theory_General}

\begin{figure*}[h]
    \centering
    \includegraphics{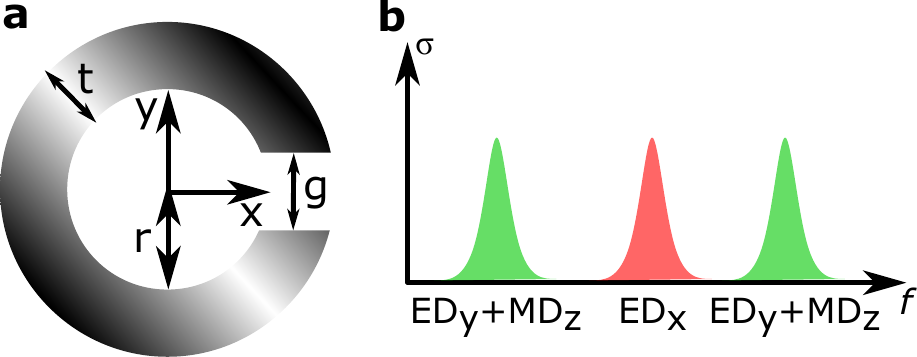}
    \caption{Bianisotropic meta-atom under study. (a) Particle geometry and parameters ($r = 2.25$ mm, $t=0.75$ mm, $g=0.75$ mm). (b) Three lowest-frequency dipole eigenmodes and their multipole content.}
    \label{fig:S1}
\end{figure*}

To demonstrate orientation-dependent coupling of bianisotropic meta-atoms, we first examine the eigenmodes of a single particle with broken mirror symmetry in the $yz$-plane (Fig.~\ref{fig:S1}a), where $z$ is the symmetry axis of an unperturbed inversion-symmetric meta-atom.

The equations governing the dipole response of such a particle read:
\begin{equation}\label{eq:bianisotropy}
\begin{gathered}
\textbf{p} = \hat{\alpha}^{\rm{ee}} \textbf{E} + \hat{\alpha}^{\rm{em}} \textbf{H},\\	
\textbf{m} = \hat{\alpha}^{\rm{me}} \textbf{E} + \hat{\alpha}^{\rm{mm}} \textbf{H},
\end{gathered}
\end{equation}
where $\mathbf{p}, \ \mathbf{m}$ are the induced electric and magnetic dipole moments, $\mathbf{E}, \ \mathbf{H}$ are external electric and magnetic fields,  $\hat{\alpha}^{\rm{ee}}$, $\hat{\alpha}^{\rm{mm}}$ are the conventional electric and magnetic polarizability tensors, and $\hat{\alpha}^{\rm{em}}$ and $\alpha^{\rm{me}}$ are $3\times 3$ magneto-electric polarizabilities (pseudo-tensors) describing the coupling of magnetic field ${\bf H}$ to the electric dipole moment ${\bf p}$ and electric field ${\bf E}$ to the magnetic dipole moment ${\bf m}$, respectively.

To retrieve the dipole eigenmodes of the meta-atom, we rewrite Eqs.~(\ref{eq:bianisotropy}) in terms of the inverse polarizability tensor and assume no external fields:
\begin{equation}\label{eq:eigenproblem}
\begin{pmatrix}
\hat{\alpha}^{\rm{ee}} & \hat{\alpha}^{\rm{em}}\\
\hat{\alpha}^{\rm{me}} & \hat{\alpha}^{\rm{mm}}
\end{pmatrix}^{-1}
\begin{pmatrix}
\textbf{p}\\
\textbf{m}
\end{pmatrix}
\equiv
\begin{pmatrix}
\hat{\zeta}^{\rm{ee}} & \hat{\zeta}^{\rm{em}}\\
\hat{\zeta}^{\rm{me}} & \hat{\zeta}^{\rm{mm}}
\end{pmatrix}
\begin{pmatrix}
\textbf{p}\\
\textbf{m}
\end{pmatrix}
= \hat{\zeta} 
\begin{pmatrix}
\textbf{p}\\
\textbf{m}
\end{pmatrix}
= 0.
\end{equation}
Symmetries of the particle which include $Oxy$ and $Oxz$ reflection planes impose constraints on the structure of $\hat{\zeta}$ tensor so that it includes only several  independent components. In particular, $\hat{\zeta}^{\rm{em}, \rm{me}}$ have only $\zeta^{\rm{em},\rm{me}}_{yz}$ and $\zeta^{\rm{em},\rm{me}}_{zy}$ nonzero components. Due to time-reversal symmetry of the particle, these components are purely imaginary $\zeta^{\rm{em}}_{yz} = iv_1$, $\zeta^{em}_{\rm{zy}} = -iv_2$, where $v_1$ and $v_2$ are real shape-dependent parameters. In addition, if the meta-atom is lossless, $(\hat{\zeta}^{\rm{em}})^{\dag} = \hat{\zeta}^{\rm{me}}$~\cite{Belov}. 

In turn, the similar symmetry analysis for the tensors $\hat{\zeta}^{\rm{ee}, \rm{mm}}$ yields that they are diagonal:
\begin{equation}\label{eq:zetaeemm}
\hat{\zeta}^{\rm{ee,mm}} = 
\begin{pmatrix}
\zeta^{e,m}_{xx} & 0 & 0 \\
0 & \zeta^{e,m}_{yy} & 0 \\
0 & 0 & \zeta^{e,m}_{zz}
\end{pmatrix}\:.
\end{equation}
In reality, all components of polarizability tensors depend on frequency. In our analysis, however, we are interested only in a sufficiently narrow spectral range that includes electric dipole resonance for in-plane dipole and magnetic dipole resonance for out-of-plane dipole at frequencies $f^e$ and $f^m$, respectively. To accommodate these assumptions, we approximate the components of the inverse polarizability by the following expressions: $\zeta_{xx}^{ee,mm} = A^{e,m}_{x}(f-f^{e,m}_x)$, $\zeta_{yy}^{ee,mm} = A^{e,m}_{y}(f-f^{e,m}_y)$ and $\zeta_{zz}^{ee,mm} = A^{e,m}_{z}(f-f^{e,m}_z)$. These expressions together result in an inverse polarizability tensor
\begin{equation}\label{eq:zeta}
\hat{\zeta}=
\begin{pmatrix}
A^e_x (f-f^e_x) & 0 & 0 & 0 & 0 & 0 \\
0 & A^e_y (f-f^e_y) & 0 & 0 & 0 & iv_1 \\
0 & 0 & A^e_z (f-f^e_z) & 0 & -iv_2 & 0 \\
0 & 0 & 0 & A^m_x (f-f^m_x) & 0 & 0 \\
0 & 0 & iv_2 & 0 & A^m_y (f-f^m_y) & 0 \\
0 & -iv_1 & 0 & 0 & 0 & A^m_z (f-f^m_z) \\
\end{pmatrix}\:,
\end{equation}
where $v_1$ and $v_2$ quantify the strength of magneto-electric hybridization in the meta-atom.

Direct solution to the eigenvalue problem (\ref{eq:eigenproblem}) with $\hat{\zeta}$ defined by Eq.~(\ref{eq:zeta}) yields six eigenmodes with the frequencies $f^e_x$, $f^m_x$, $f^{(1)}_{\pm} = \dfrac{(f^e_z + f^m_y)}{2} \pm \sqrt{\dfrac{(f^e_z-f^m_y)^2}{4} + \dfrac{v_2^2}{A^e_z A^m_y}}$ and $f^{(2)}_{\pm} = \dfrac{(f^e_y + f^m_z)}{2} \pm \sqrt{\dfrac{(f^e_y-f^m_z)^2}{4} + \dfrac{v_1^2}{A^e_y A^m_z}}$. Inspecting the associated eigenvectors, we recover that the modes with frequencies $f^e_x$ and $f^m_x$ contain only $x$-component of the electric and magnetic dipole moments, respectively, whereas the modes with the frequencies $f^{(1)}_{\pm}$ ($f^{(2)}_{\pm}$) exhibit hybrid nature having both $z$ ($y$)-oriented electric dipole moment and $y$ ($z$)-oriented magnetic dipole moment. 

In this work, we focus on the lower-frequency hybrid dipole mode $f^{(2)}_{-} = f_{-}$ having nonzero $p_y$ and $m_z$ components of the dipole moments, which is well-separated from the rest of the modes. 

To simplify the notations, we reassign the eigenfrequency of dipole resonances as $f^e_y = f_e$ and $f^m_z = f_m$ and the associated coefficients $A^e_y = A_e$ and $A^m_z = B_m$. In the vicinity of these two resonances, frequency dependence of the remaining components of polarizability tensor can be neglected and hence the inverse polarizability tensor takes the following approximate form:
\begin{equation}\label{eq:zeta_simple}
\hat{\zeta} = 
\begin{pmatrix}
A_e (f-f_e) & 0 & 0 & 0 & 0 & 0 \\
0 & A_e (f-f_e) & 0 & 0 & 0 & iv_1 \\
0 & 0 & B_e & 0 & -iv_2 & 0 \\
0 & 0 & 0 & A_m & 0 & 0 \\
0 & 0 & iv_2 & 0 & A_m & 0 \\
0 & -iv_1 & 0 & 0 & 0 & B_m (f-f_m) \\
\end{pmatrix}
\end{equation}

\begin{figure*}[h]
    \centering
    \includegraphics{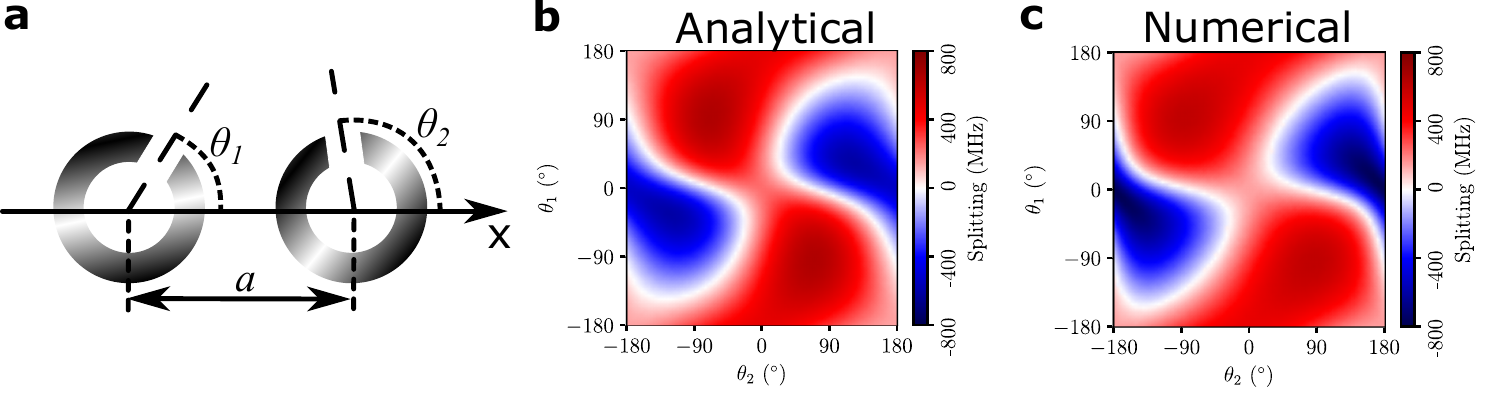}
    \caption{Analysis of orientation-dependent coupling of the two bianisotropic meta-atoms designed as split-ring resonators (SRRs). (a) Geometry of the considered dimer with the distance between the SRRs $a = 7.5$ mm and rotation angles $\theta_1$ and $\theta_2$. (b, c) Analytical and numerical results for the splitting of the lower-frequency hybrid dipole mode due to interaction of the two SRRs.}
    \label{fig:S2}
\end{figure*}

Using the developed model, we investigate the interaction of two bianisotropic meta-atoms depending on their mutual orientation described by the two angles of their rotation (Fig.~\ref{fig:S2}a). As a specific implementation of the meta-atom, we choose split-ring resonator (SRR).

Rotation of the SRR transforms its inverse polarizability tensor according to the law $\hat{\zeta}' = \hat{U}_{1,2} \hat{\zeta} \hat{U}_{1,2}^{-1}$, where unitary rotation operator $U_{1,2}$ is given by:
\begin{equation}
\hat{U}_{1,2}
= \begin{pmatrix}
\cos\theta_{1,2} & \sin\theta_{1,2} & 0\\
-\sin\theta_{1,2} & \cos\theta_{1,2} & 0\\
0 & 0 & 1
\end{pmatrix}.
\end{equation}
The inverse polarizability tensor of a single bianisotropic meta-atom rotated by the angle $\theta_{1,2}$ reads:
\begin{equation}
\hat{\zeta}(\theta_{1,2}) = 
\begin{pmatrix}
A_e (f-f_e) & 0 & 0 & 0 & 0 & iv_1\sin\theta_{1,2} \\
0 & A_e (f-f_e) & 0 & 0 & 0 & iv_1\cos\theta_{1,2} \\
0 & 0 & B_e & -iv_2\sin\theta_{1,2} & -iv_2\cos\theta_{1,2} & 0 \\
0 & 0 & iv_2\sin\theta_{1,2} & A_m & 0 & 0 \\
0 & 0 & iv_2\cos\theta_{1,2} & 0 & A_m & 0 \\
-iv_1\sin\theta_{1,2} & -iv_1\cos\theta_{1,2} & 0 & 0 & 0 & B_m (f-f_m) \\
\end{pmatrix}
\end{equation}

Employing the dipole model, we describe the interaction of the two bianisotropic meta-atoms (Fig.~\ref{fig:S2}a) by the equations:
\begin{equation}
\begin{gathered}
\hat{\zeta}(\theta_{1}) \begin{pmatrix}
\textbf{p}_1\\
\textbf{m}_1
\end{pmatrix} =
\begin{pmatrix}
G^{\rm{ee}}(\mathbf{r}_1-\mathbf{r}_2) & G^{\rm{em}}(\mathbf{r}_1-\mathbf{r}_2)\\
G^{\rm{me}}(\mathbf{r}_1-\mathbf{r}_2) & G^{\rm{mm}}(\mathbf{r}_1-\mathbf{r}_2)
\end{pmatrix} \begin{pmatrix}
\textbf{p}_2\\
\textbf{m}_2
\end{pmatrix} = \hat{G}_{12} \begin{pmatrix}
\textbf{p}_2\\
\textbf{m}_2
\end{pmatrix},\\
\hat{\zeta}(\theta_{2}) \begin{pmatrix}
\textbf{p}_2\\
\textbf{m}_2
\end{pmatrix} =
\begin{pmatrix}
G^{\rm{ee}}(\mathbf{r}_2-\mathbf{r}_1) & G^{\rm{em}}(\mathbf{r}_2-\mathbf{r}_1)\\
G^{\rm{me}}(\mathbf{r}_2-\mathbf{r}_1) & G^{\rm{mm}}(\mathbf{r}_2-\mathbf{r}_1)
\end{pmatrix} \begin{pmatrix}
\textbf{p}_1\\
\textbf{m}_1
\end{pmatrix} = \hat{G}_{21} \begin{pmatrix}
\textbf{p}_1\\
\textbf{m}_1
\end{pmatrix}.
\end{gathered}
\end{equation}
Here, $G^{\rm{ee},\rm{mm}}$ denote Green's dyadics, responsible for the electric (magnetic) fields produced by electric (magnetic) dipoles. In turn, $G^{\rm{em,me}}$ are related to the ``cross-fields'', i.e. the electric (magnetic) fields produced by magnetic (electric) dipoles, which are absent in the quasistatic limit\cite{Novotny}. However, accounting for them turns out to be crucial to track angle-dependent coupling. In our analysis, we expand the Green's functions with respect to $\omega\,r/c$ keeping the terms up to the first order, which yields:
\begin{equation}
G^{\rm{ee}}(|\mathbf{r}_1-\mathbf{r}_2|) = G^{\rm{mm}}(|\mathbf{r}_1-\mathbf{r}_2|) =
\begin{pmatrix}
2\lambda & 0 & 0 \\
0 & -\lambda & 0 \\
0 & 0 & -\lambda
\end{pmatrix},
\end{equation}
\begin{equation}
G^{\rm{em}}(|\mathbf{r}_1-\mathbf{r}_2|) = -G^{\rm{me}}(|\mathbf{r}_1-\mathbf{r}_2|) =
\begin{pmatrix}
0 & 0 & 0 \\
0 & 0 & -i\lambda qa \\
0 & i\lambda qa & 0
\end{pmatrix},
\end{equation}
where $\lambda = 1/a^3$, $a$ is the distance between the particles and $q = \omega/c$. To obtain tractable expressions for the eigenfrequency splittings, we treat particle's interaction as a perturbation keeping only the leading-order terms in small parameter $\lambda$. 

The described assumptions result in the following eigenvalue problem:
\begin{equation}\label{eq:dimer_eigs}
\begin{vmatrix}
\hat{\alpha}^{-1}(\theta_1) & -G_{12}\\
-G_{21} & \hat{\alpha}^{-1}(\theta_2)
\end{vmatrix} = 0,
\end{equation}
where off-diagonal elements are proportional to the small parameter $\lambda$ and are considered as a perturbation. In the leading order in $\lambda$, we recover the following results: $f_e$ splits into two new frequencies $f_e \pm \Delta_e/2$, where $\Delta_e(\theta) = \dfrac{3\cos(\theta_1+\theta_2) + \cos(\theta_1-\theta_2)}{A_e a^3}$, whereas the frequencies $f_{\pm}$ associated with hybrid magneto-electric modes acquire the splittings
$\Delta_{\pm}~=~A_{\pm}+B_{\pm}(\cos\theta_1-\cos\theta_2)+C_{\pm}(\cos\theta_1\cos\theta_2-2\sin\theta_1\sin\theta_2),$ where $A_{\pm} = \frac{2 (f_{1,2} - f_e)}{a^3 B_m (f_1-f_2)}$, $B_{\pm} = \frac{4 \pi f_{1,2}}{v_1\,c a^2 A_e B_m (f_1-f_2)}$ and $C_{\pm} = \frac{2 v_1^2}{a^3 A_e^2 B_m (f_{1,2}-f_e)(f_1-f_2)}$. Since we focus on the lower-frequency hybrid dipole mode with the frequency $f_{-}$, from now on we suppress the subscript ``-'':
\begin{equation}\label{eq:generalized_dimer_splitting}
    \Delta(\theta_1, \theta_2) = A + B(\cos\theta_1-\cos\theta_2) + C(\cos\theta_1\cos\theta_2-2\sin\theta_1\sin\theta_2).
\end{equation}

To validate the elaborated theoretical model for the angle-dependent coupling, we simulate the eigenfrequency splittings using eigenfrequency solver of Comsol Multiphysics software package, wave optics module. The obtained data is then fitted by Eq.~(\ref{eq:generalized_dimer_splitting}) which captures the numerical data  qualitatively correctly (Fig.~\ref{fig:S2}b,c).  Note that the consistent check would require the knowledge of all coefficients $A_e, \ A_m, \ B_e, \ B_m$, retrieved independently, for instance, from the scattering spectra. In this case, however, we are interested only in a qualitative agreement.

Besides the strong dependence on the angles $\theta_1, \theta_2$, it is also noteworthy that the splitting $\Delta = f_a - f_s$ changes the sign for some angles indicating that the symmetric ($f_s$) and antisymmetric ($f_a$) eigenmodes swap their places. It should be stressed that achieving the same functionality in the tight-binding lattices is more challenging since it requires auxiliary off-resonant sites to change the sign of the coupling constant\cite{Keil}.

\begin{figure*}[h!]
    \centering
    \includegraphics[width=\linewidth]{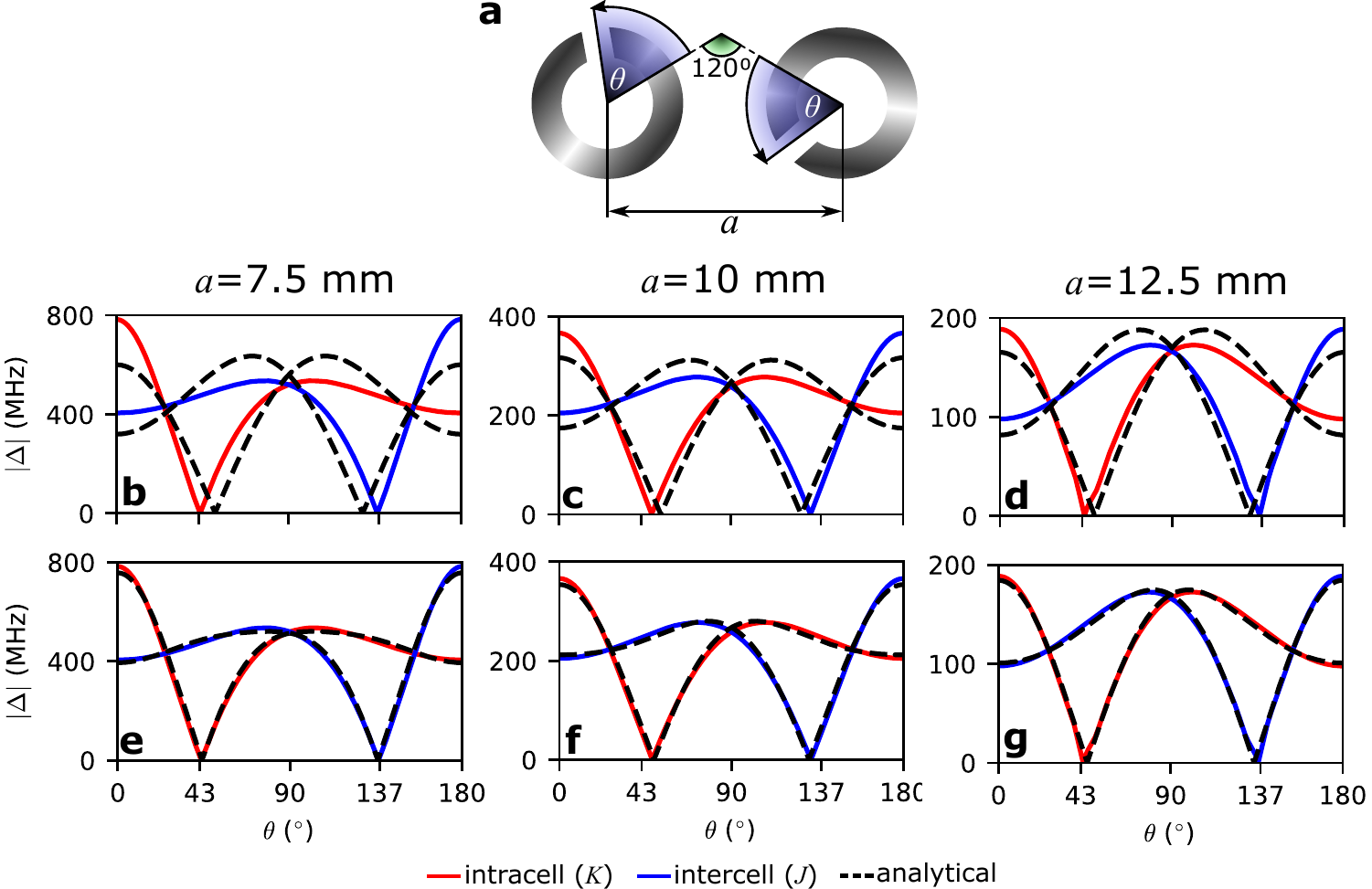}
    \caption{Splitting of the lower-frequency hybrid dipole mode due to the interaction between the two bianisotropic meta-atoms versus rotation angle $\theta$ for the geometry depicted in panel (a). (b-d) Comparison of the results of full-wave numerical simulations with the derived expression Eq.~\eqref{eq:dimer_splitting} based on the dipole model. (e-g) Fit of the numerical results by the series Eq.~\eqref{eq:series}. Analytical curve on the panel (e) contains $A_0-A_4$ terms ($A_0 = -210.9$ MHz, $A_1 = 453.6$ MHz, $A_2 = 347.8$ MHz, $A_3 = 122.4$ MHz and $A_4 = 44.7$ MHz), whereas panels  (f,g) include only the first four terms. (f) $A_0=-98.2$ MHz, $A_1=243.4$ MHz, $A_2=169.2$ MHz, $A_3=39.6$ MHz. (g) $A_0=-64.1$ MHz, $A_1=122.3$ MHz, $A_2=105.9$ MHz, $A_3=20.8$ MHz.}
    \label{fig:S3}
\end{figure*}

Given that the negative couplings are required for quadrupole topological phase engineering, our approach may open an access to tunable quadrupole topological insulators. In this work, however, we focus on higher-order topological states protected by $C_3$ symmetry. To preserve this symmetry at the level of the lattice, angles $\theta_{1,2}$, should be linked to each other as $\theta_1 = \pi/6 + \theta$ and $\theta_2 = 5\pi/6 + \theta$, where $\theta$ can be chosen arbitrarily as illustrated in Fig.~\ref{fig:S3}a. Then the expression for the splitting of the lower frequency hybrid mode, Eq.~(\ref{eq:generalized_dimer_splitting}) reduces to
\begin{equation}\label{eq:dimer_splitting}
\Delta = \Tilde{A} + \Tilde{B}\cos\theta + \Tilde{C}\cos2\theta,
\end{equation}
where $\Tilde{A} = A + \frac{C}{2}$, $\Tilde{B} = B\sqrt{3}$ and $\Tilde{C} = -\frac{3}{2}C$. Due to the term $\propto\cos\theta$, the ratio between the effective couplings of the two particles with rotation angles $(\theta, \ \pi-\theta)$ and $(\pi-\theta, \ \theta)$ $\xi = \Delta(\theta)/\Delta(\pi-\theta) \ne 1$ which allows us to control the ratio of the meta-atom coupling constants via the rotation angle $\theta$. It should be emphasized that the term proportional to $\cos\theta$ in Eq.~(\ref{eq:dimer_splitting}) vanishes in the absence of bianisotropy or if the magneto-electric part of the Green's function is neglected.

To probe the validity of Eq.~(\ref{eq:dimer_splitting}) additionally, we perform detailed numerical simulations of eigenfrequency splittings in the SRR pair (Fig.\ref{fig:S3}a) and fit the obtained numerical data by Eq.~\eqref{eq:dimer_splitting}. Interestingly, the greater the distance between the meta-atoms, the better is the agreement between the analytical curve and numerical  data~Fig.\ref{fig:S3}b-d. We attribute such behavior to the excitation of higher-order multipoles in the adjacent particles as well as to the response of the dipole moments to the field gradients\cite{Bobylev}.

To incorporate these phenomena, we present the eigenfrequency splitting as
\begin{equation}\label{eq:series}
    \Delta(\theta) = \sum_{n} A_n \cos n\theta,
\end{equation}
which is a Fourier series of an even periodic function $\Delta(\theta)$. This formula not only grasps the physics beyond the developed dipole model, Eq.~\eqref{eq:dimer_splitting}, but also provides an increased accuracy in the description of orientation-dependent coupling (Fig.~\ref{fig:S3}e-g) compared to the dipole model, Eq.~(\ref{eq:dimer_splitting}). In this case, harmonics with odd $n$ provide the tool to engineer staggered effective coupling patterns to create analog of breathing kagome lattice\cite{Ezawa2018}, when the intercell coupling is greater than the intracell one (Fig.~\ref{fig:S4}).

\newpage

\section{Supplementary Note 2 -- Calculation of topological invariant}\label{sec:invariant}

\begin{figure*}[h]
    \centering
    \includegraphics[width=\linewidth]{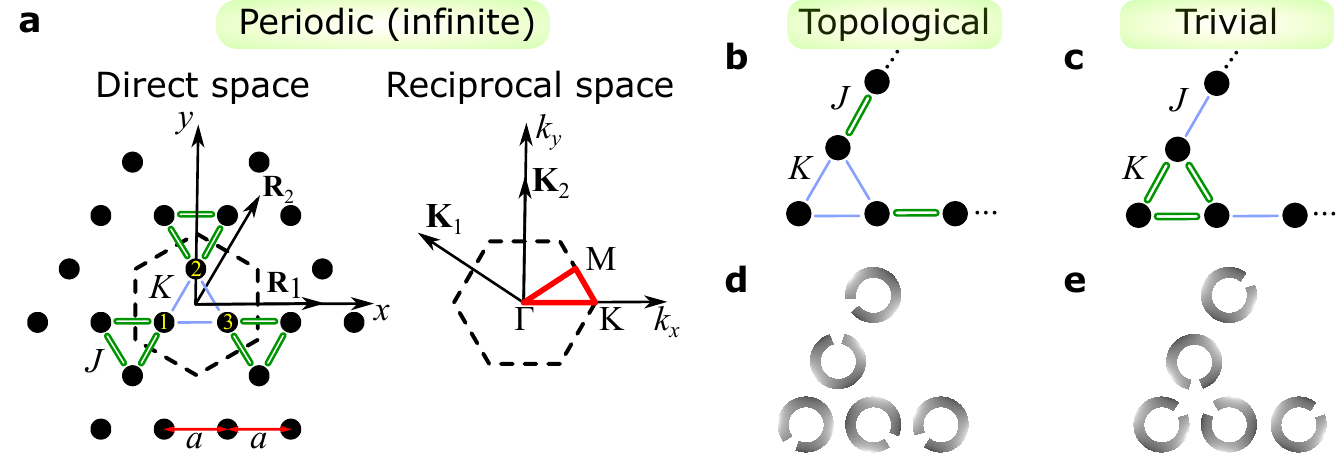}
    \caption{Scheme of kagome lattice based on rotated SRRs. (a) Lattices in the real and reciprocal space. Given the distance $a$ between the neighboring sites, primitive translation vectors and the associated reciprocal vectors read $\mathbf{R}_1 = (2a, 0)$, $\mathbf{R}_2 = (a, a\sqrt{3})$ and $\mathbf{K}_1 = (\pi/a, -\pi/(a\sqrt{3}))$, $\mathbf{K}_2 = (0, 2\pi/(a\sqrt{3}))$. (b, c) Corners of topological and trivial structures in the tight-binding model and (d, e) their realizations with the help of split-ring resonators.}
    \label{fig:S4}
\end{figure*}

To prove the topological origin of our model, we calculate the relevant topological invariant for $C_{3}$-symmetric system\cite{2019_Benalcazar}:
\begin{equation}
\chi^{(3)} = ([K_1^{(3)}], [K_2^{(3)}])
\end{equation}
where $[K_{1,2}^{(3)}] = \# K_{1,2}^{(3)} - \# \Gamma_{1,2}^{(3)}$. Here, $K$ and $\Gamma$ denote high-symmetry points of the first Brillouin zone (Fig.~\ref{fig:S4}a), the superscript stands for the 3-fold symmetry of the system and the subscript $p$ indicates the type of the eigenstate rotational symmetry:  the eigenstates multiply by the factor $e^{2\pi(p-1)/3}$ upon $2\pi/3$ rotation, where $p$ can be equal to 1, 2 or 3. The \# sign preceding each of the two terms denotes the number of bands with a given symmetry below the gap under consideration. In particular, we consider the gap  between the bottom and the middle bands. 

We retrieve Bloch modes directly from the full-wave numerical simulation of a periodic structure with different angles of SRRs rotation preserving $C_3$ symmetry of the structure. Due to the symmetry of the problem, the obtained Bloch vectors transform under rotation by $2\pi/3$:
\begin{equation}
\hat{R} \ket{\psi} = e^{2\pi(p-1)/3}\,\ket{\psi}\:.
\end{equation}
where $\hat{R}$ is a rotation matrix by the angle $2\pi/3$. For simplicity, we assume that the Bloch modes are 3-component vectors comprised only of the $z$-components of magnetic dipole moments, so that the rotation results in the permutation of these components. Simulated results for the angle $\theta=\pi$ are provided in Table 1. It is clearly seen that for $p=1$ $\# \Gamma_{1}^{(3)} = 1$ and $\# K_{1}^{(3)} = 0$, so $[K_{1}^{(3)}] = 0 - 1 = -1$. Similarly, $[K_2^{(3)}] = 0$, which yields $\chi^{(3)} = (-1, 0)$. This result indicates the non-zero topological invariant of our system (Fig.\ref{fig:S4}b-e).

\begin{table}
\begin{center}
\begin{tabular}{l l l c c}
\toprule
$\Gamma$-point & ($k_x=0$, $k_y=0$): & & \\
\toprule
 & $\centering{f \ (\rm{GHz})} $  & $\centering{\psi \sim (m_{z1}, m_{z2}, m_{z3})}$ & $\centering{p}$ \\
 \cmidrule(r){2-5}
 & $f_1= 0$   &  $\psi_1=(1,1,1)$ & $1$ \\
 & $f_2= 7.118$  &  $\psi_2=(1,e^{i\frac{2\pi}{3}},e^{i\frac{4\pi}{3}})$ & $2$ \\
 & $f_3= 7.120$  &  $\psi_3=(1,e^{i\frac{4\pi}{3}},e^{i\frac{2\pi}{3}})$ & $3$ \\
 \toprule
${\rm K}$-point & ($k_x=\pi/(a\sqrt{3})$, $k_y=0$) & & \\
 \toprule
 \cmidrule(r){2-4}
 & $f_1= 6.012$  &  $\psi_1=(1,e^{i\frac{4\pi}{3}},e^{i\frac{2\pi}{3}})$ & $3$ &\\
 & $f_2= 7.185$  &  $\psi_2=(1,e^{i\frac{2\pi}{3}},e^{i\frac{4\pi}{3}})$   & $2$ &\\
 & $f_3= 7.310$  &  $\psi_3=(1,1,1)$ & $1$ &\\
 \toprule
\end{tabular}
\caption{\label{tab:Topological} Calculation of the topological invariant for topologically nontrivial case with SRR rotation angle $\theta=\pi$. $f_{i}$ and $\psi_{i}$ are  eigenvalues and Bloch eigenfunctions retrieved from numerical simulations for $i=1,2,3$. Index $i$ enumerates the bands from the bottom.  $p$ is a symmetry index of eigenvalues of the  rotation operator $\hat{R}$ by the angle $2\,\pi/3$ given by the expressions $\exp(2\pi(p-1) i/3)$. Photonic bandgap occupies the frequency range from approximately 6.0~GHz to 6.8~GHz.}
\end{center}
\end{table}

\begin{figure*}[h]
    \centering
    \includegraphics[width=0.6\linewidth]{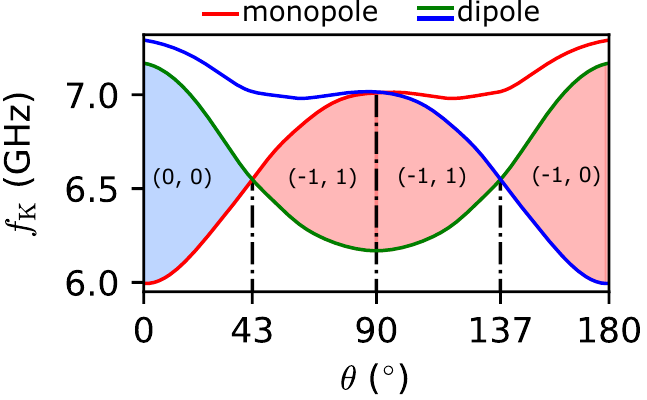}
    \caption{The frequencies of the lowest-frequency dipole modes in K point of the Brillouin zone versus rotation angle $\theta$. The indices in brackets display the topological invariant $\chi^{(3)}$ for the different values of the rotation angle.}
    \label{fig:S5}
\end{figure*}

The calculated values of the invariant for the other rotation angles $\theta$ from $0^\circ$ to $180^\circ$ are illustrated in Fig.~\ref{fig:S5}. Based on the obtained topological indices, we determine the topological bulk polarization for each of the parameter domains in~Fig.\ref{fig:S5}:
\begin{equation*}
\begin{gathered}
0-43^{\circ}: \ \mathbf{P} = 0,\\
43-90^{\circ}: \ \mathbf{P} = \frac{2e}{3}(\mathbf{R}_1 + \mathbf{R}_2),\\
90-137^{\circ}: \ \mathbf{P} = \frac{2e}{3}(\mathbf{R}_1 + \mathbf{R}_2),\\
137-180^{\circ}: \ \mathbf{P} = -\frac{2e}{3}(\mathbf{R}_1 + \mathbf{R}_2),
\end{gathered}
\end{equation*}
where $\mathbf{R}_{1,2}$ are the translation vectors (Fig.~\ref{fig:S4}a).

\newpage

\newpage

\section{Supplementary Note 3 -- Calculation of the local density of states}
\label{sec:ldos}

In our experiments, we have measured the full map of $S_{12}$ coefficients with the spatial resolution 1~mm along both axes in the frequency range from 5.8 to 7.2 GHz. A natural way to visualize the results is provided by the concept of local density of states, which is defined as a sum
\begin{equation}\label{LDOS1}
    LDOS=\sum\limits_{m,n}\,|S_{12}^{(mn)}|^2\,h^{(mn)}\:,
\end{equation}
where $S_{12}^{(mn)}$ is a complex transmission coefficient when the receiving antenna is placed near $(m,n)$ pixel and $h^{(mn)}$ is a spatial filter function equal either to 1 or to 0 depending on the pixel.

The key idea behind the choice of the spatial filter function is to enhance the contribution from the specific type of modes while suppressing the contribution from the others.

\begin{figure*}[h]
    \centering
    \includegraphics[width=\linewidth]{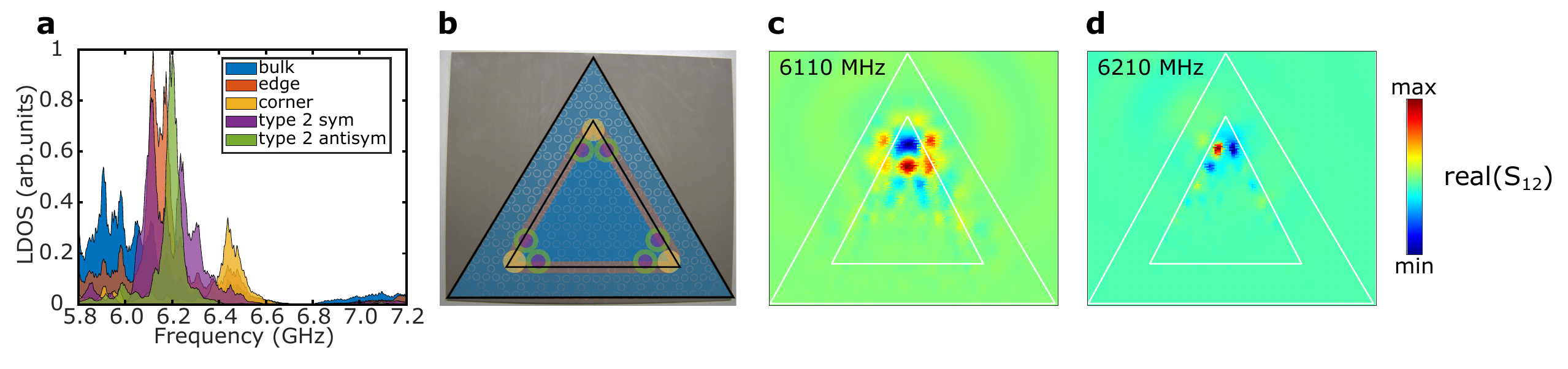}
    \caption{Experimental investigation of topological edge and corner states. Period of kagome lattice is equal to $a=15$~mm, inner radius of the ring $r=2.25$~mm, ring width and ring gap $t=b=0.75$~mm, the distance between the neighboring rings is $a=7.5$ mm. (a) Retrieved local density of states at the corner, edge and bulk of the structure. Additional post-processing of the data allows us to reveal type-II corner states. (b) Photograph of experimental sample with overlaid spatial filters used to extract different types of the local density of states. (c, d) Field profiles of symmetric and antisymmetric type-II corner states.}
    \label{fig:S6}
\end{figure*}

First we consider the contribution from the bulk sites setting $h^{(mn)}=1$ in all bulk sites [see blue domain in Fig.~\ref{fig:S6}b] and $h^{(mn)}=0$ at the edge and at the corner. The obtained result depicted in blue in Fig.~\ref{fig:S6}a shows a series of peaks forming a continuum that corresponds to the bulk band. Higher-frequency bulk band is excited relatively weakly which is due to the chosen position of exciting dipole antenna  close to the corner of the structure.

Next we consider the contribution from the edge sites setting the spatial filter function to be nonzero only at the edge sites [see red domain in Fig.~\ref{fig:S6}b]. This yields the red curve in Fig.~\ref{fig:S6}a which indicates the spectral position of the band of edge states. The characteristic field profile of such edge-localized states is illustrated in Fig.~2 of the main text.

Finally, we focus on the corner states arising due to the higher-order topology of the studied system. To capture such modes, we select spatial filter function to be nonzero only in the vicinity of the corner which is expected to yield maximum at the frequency of  corner-localized modes. In agreement with this intuition, orange curve in Fig.~\ref{fig:S6}a features a pronounced maximum with the associated field distribution depicted in Fig.~2 of the main text. The observed edge state corresponds to that expected in the nearest-neighbor coupled kagome lattice.

However, due to the long-range nature of electromagnetic interactions two more corner modes are expected to emerge splitting from the band of edge states\cite{Zhirihin}. To isolate the associated peaks in the experimental data, the technique of LDOS calculation has to be refined in such a way, that is would be sensitive only to the modes with the specific symmetry: either  symmetric or antisymmetric relative to reflection in the bisectrix of the respective corner. Thus, we construct the quantity
\begin{equation}\label{LDOS2}
    LDOS_{\pm}=\sum\limits_{m,n}\,|S_{12}^{(mn)}\pm S_{12}^{(nm)}|^2\,h^{(mn)}\:,
\end{equation}
where $(m,n)$ and $(n,m)$ pixels are assumed to be mirror-symmetric with respect to the bisectrix of the corner, while $h^{(mn)}=h^{(nm)}=1$.

Frequency dependencies of LDOS$_{\pm}$ depicted in Fig.~\ref{fig:S6}{\bf a} exhibit two sharp peaks with the associated field distribution depicted in Fig.~\ref{fig:S6}c,d.  These distributions reveal corner-localized modes with symmetric (LDOS$_{+}$) and antisymmetric (LDOS$_{-}$) field distributions. Both of them closely resemble the modes expected in the systems with long-range interactions of the meta-atoms.